\documentclass[article,nojss]{jss}
\usepackage[utf8]{inputenc}

\providecommand{\tightlist}{%
  \setlength{\itemsep}{0pt}\setlength{\parskip}{0pt}}

\author{
Iñaki Ucar\\Universidad Carlos III de Madrid \And Bart Smeets\\dataroots \And Arturo Azcorra\\Universidad Carlos III de Madrid\\
IMDEA Networks Institute
}
\title{\pkg{simmer}: Discrete-Event Simulation for \proglang{R}}

\Plainauthor{Iñaki Ucar, Bart Smeets, Arturo Azcorra}
\Plaintitle{simmer: Discrete-Event Simulation for R}

\Abstract{
The \pkg{simmer} package brings discrete-event simulation to
\proglang{R}. It is designed as a generic yet powerful process-oriented
framework. The architecture encloses a robust and fast simulation core
written in \proglang{C++} with automatic monitoring capabilities. It
provides a rich and flexible \proglang{R} API that revolves around the
concept of \emph{trajectory}, a common path in the simulation model for
entities of the same type.
}

\Keywords{\proglang{R}, Discrete-Event Simulation}
\Plainkeywords{R, Discrete-Event Simulation}

\Address{
    Iñaki Ucar\\
  Universidad Carlos III de Madrid\\
  Avda. de la Universidad, 30 28911 Leganés (Madrid) Spain\\
  E-mail: \href{mailto:inaki.ucar@uc3m.es}{\nolinkurl{inaki.ucar@uc3m.es}}\\

      Bart Smeets\\
  dataroots\\
  Witte Patersstraat, 4 1040 Brussels Belgium\\
  E-mail: \href{mailto:bart@dataroots.io}{\nolinkurl{bart@dataroots.io}}\\

      Arturo Azcorra\\
  Universidad Carlos III de Madrid\\
  IMDEA Networks Institute\\
  Avda. de la Universidad, 30 28911 Leganés (Madrid) Spain\\
  E-mail: \href{mailto:azcorra@it.uc3m.es}{\nolinkurl{azcorra@it.uc3m.es}}\\

  }

\usepackage{amsmath} \usepackage{longtable} \usepackage{booktabs}
\usepackage{subfig}

\begin{document}

\setlength{\abovedisplayskip}{0pt}

This manuscript corresponds to \pkg{simmer} version 3.6.4 and was
typeset on December 05, 2017. For citations, please use the version
accepted on the JSS when available.

\section{Introduction}\label{introduction}

The complexity of many real-world systems involves unaffordable
analytical models, and consequently, such systems are commonly studied
by means of simulation. This problem-solving technique precedes the
emergence of computers, but tool and technique got entangled as a result
of their development. As defined by \citet{Shannon:1975:Systems},
simulation ``is the process of designing a model of a real system and
conducting experiments with this model for the purpose either of
understanding the behaviour of the system or of evaluating various
strategies (within the limits imposed by a criterion or a set of
criteria) for the operation of the system.''

Different types of simulation apply depending on the nature of the
system under consideration. A common model taxonomy classifies
simulation problems along three main dimensions
\citep{Law:2000:Simulation}: (\emph{i}) deterministic vs.~stochastic,
(\emph{ii}) static vs.~dynamic (depending on whether they require a
\emph{time} component), and (\emph{iii}) continuous vs.~discrete
(depending on how the system changes). For instance, \emph{Monte Carlo
methods} are well-known examples of static stochastic simulation
techniques. On the other hand, \emph{Discrete-event simulation} (DES) is
a specific technique for modelling stochastic, dynamic and discretely
evolving systems. As opposed to \emph{continuous simulation}, which
typically uses smoothly-evolving equational models, DES is characterised
by sudden state changes at precise points of (simulated) time.

Customers arriving at a bank, products being manipulated in a supply
chain, or packets traversing a network are common examples of such
systems. The discrete nature of a given system arises as soon as its
behaviour can be described in terms of \emph{events}, which is the most
fundamental concept in DES. An \emph{event} is an instantaneous
occurrence that may change the state of the system, while, between
events, all the state variables remain constant.

The applications of DES are vast, including, but not limited to, areas
such as manufacturing systems, construction engineering, project
management, logistics, transportation systems, business processes,
healthcare and telecommunications networks \citep{Banks:2005:Discrete}.
The simulation of such systems provides insights into the process' risk,
efficiency and effectiveness. Also, by simulation of an alternative
configuration, one can proactively estimate the effects of changes to
the system. In turn, this allows one to get clear insights into the
benefits of process redesign strategies (e.g., extra resources). A wide
range of practical applications is prompted by this, such as analysing
bottlenecks in customer services centres, optimising patient flows in
hospitals, testing the robustness of a supply chain or predicting the
performance of a new protocol or configuration of a telecommunications
network.

There are several world views, or programming styles, for DES
\citep{Banks:2005:Discrete}. In the \emph{activity-oriented} approach, a
model consists of sequences of activities, or operations, waiting to be
executed depending on some conditions. The simulation clock advances in
fixed time increments. At each step, the whole list of activities is
scanned, and their conditions, verified. Despite its simplicity, the
simulation performance is too sensitive to the election of such a time
increment. Instead, the \emph{event-oriented} approach completely
bypasses this issue by maintaining a list of scheduled events ordered by
time of occurrence. Then, the simulation just consists in jumping from
event to event, sequentially executing the associated routines. Finally,
the \emph{process-oriented} approach refines the latter with the
addition of interacting \emph{processes}, whose activation is triggered
by events. In this case, the modeller defines a set of processes, which
correspond to entities or objects of the real system, and their life
cycle.

\pkg{simmer} \citep{CRAN:simmer} is a DES package for \proglang{R} which
enables high-level process-oriented modelling, in line with other modern
simulators. But in addition, it exploits the novel concept of
\emph{trajectory}: a common path in the simulation model for entities of
the same type. In other words, a trajectory consist of a list of
standardised actions which defines the life cycle of equivalent
processes. This design pattern is flexible and simple to use, and takes
advantage of the chaining/piping workflow introduced by the
\pkg{magrittr} package \citep{CRAN:magrittr}.

Let us illustrate this with a simple example taken from
\citet{Pidd:1988:Computer}, Section 5.3.1:

\begin{quote}
Consider a simple engineering job shop that consists of several
identical machines. Each machine is able to process any job and there is
a ready supply of jobs with no prospect of any shortages. Jobs are
allocated to the first available machine. The time taken to complete a
job is variable but is independent of the particular machine being used.
The machine shop is staffed by operatives who have two tasks:

\begin{enumerate}
\def\labelenumi{\arabic{enumi}.}
\tightlist
\item
  RESET machines between jobs if the cutting edges are still OK.
\item
  RETOOL those machines with cutting edges that are too worn to be
  reset.
\end{enumerate}

In addition, an operator may be AWAY while attending to personal needs.
\end{quote}

\begin{figure}
\centering
\includegraphics[width=0.80000\textwidth]{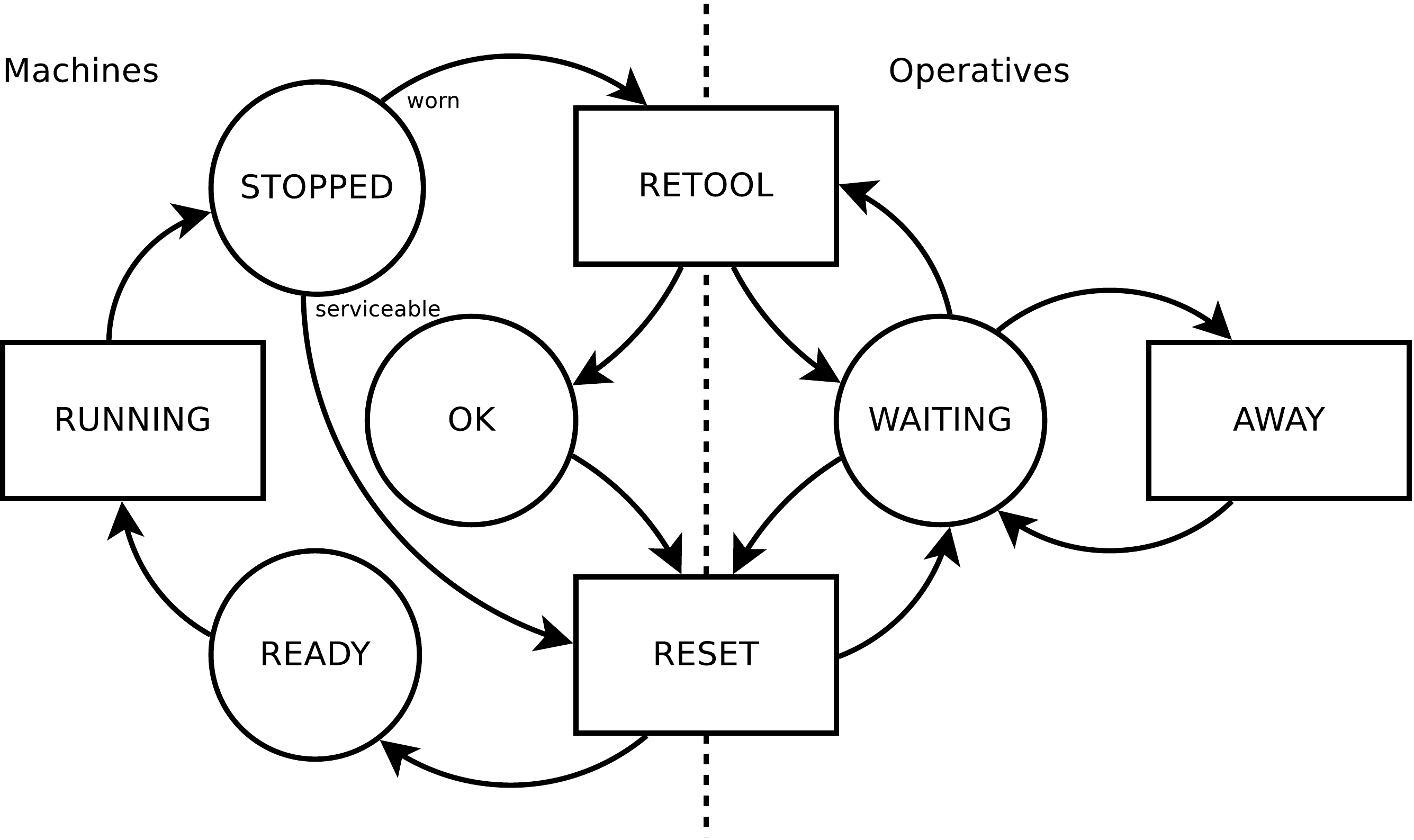}
\caption{The job shop activity cycle diagram.\label{job-shop}}
\end{figure}

Figure \ref{job-shop} shows the activity cycle diagram for the
considered system. Circles (READY, STOPPED, OK, WAITING) represent
states of the machines or the operatives respectively, while rectangles
(RUNNING, RETOOL, RESET, AWAY) represent activities that take some
(random) time to complete. Two kind of processes can be identified: shop
jobs, which use machines and degrade them, and personal tasks, which
take operatives AWAY for some time. There is a natural way of simulating
this system with \pkg{simmer} which consist in considering machines and
operatives as resources, and describing the life cycles of shop jobs and
personal tasks as trajectories.

First of all, let us instantiate a new simulation environment and define
the completion time for the different activities as random draws from
exponential distributions. Likewise, the interarrival times for jobs and
tasks are defined (\code{NEW\_JOB}, \code{NEW\_TASK}), and we consider a
probability of 0.2 for a machine to be worn after running a job
(\code{CHECK\_JOB}).

\begin{CodeInput}
R> library(simmer)
R>
R> set.seed(1234)
R>
R> (env <- simmer("Job Shop"))
\end{CodeInput}

\begin{CodeOutput}
simmer environment: Job Shop | now: 0 | next:
\end{CodeOutput}

\begin{CodeInput}
R> RUNNING <- function() rexp(1, 1)
R> RETOOL <- function() rexp(1, 2)
R> RESET <- function() rexp(1, 3)
R> AWAY <- function() rexp(1, 1)
R> CHECK_WORN <- function() runif(1) < 0.2
R> NEW_JOB <- function() rexp(1, 5)
R> NEW_TASK <- function() rexp(1, 1)
\end{CodeInput}

The trajectory of an incoming job starts by seizing a machine in READY
state. It takes some random time for RUNNING it after which the
machine's serviceability is checked. An operative and some random time
to RETOOL the machine may be needed, and either way an operative must
RESET it. Finally, the trajectory releases the machine, so that it is
READY again. On the other hand, personal tasks just seize operatives for
some time.

\begin{CodeInput}
R> job <- trajectory() %>%
R+   seize("machine") %>%
R+   timeout(RUNNING) %>%
R+   branch(
R+     CHECK_WORN, continue = TRUE,
R+     trajectory() %>%
R+       seize("operative") %>%
R+       timeout(RETOOL) %>%
R+       release("operative")
R+   ) %>%
R+   seize("operative") %>%
R+   timeout(RESET) %>%
R+   release("operative") %>%
R+   release("machine")
R>
R> task <- trajectory() %>%
R+   seize("operative") %>%
R+   timeout(AWAY) %>%
R+   release("operative")
\end{CodeInput}

Once the processes' trajectories are defined, we append 10 identical
machines and 5 operatives to the simulation environment, as well as two
generators for jobs and tasks.

\begin{CodeInput}
R> env %>%
R+   add_resource("machine", 10) %>%
R+   add_resource("operative", 5) %>%
R+   add_generator("job", job, NEW_JOB) %>%
R+   add_generator("task", task, NEW_TASK) %>%
R+   run(until=1000)
\end{CodeInput}

\begin{CodeOutput}
simmer environment: Job Shop | now: 1000 | next: 1000.09508921831
{ Resource: machine | monitored: TRUE | server status: 3(10) | queue... }
{ Resource: operative | monitored: TRUE | server status: 2(5) | queue... }
{ Generator: job | monitored: 1 | n_generated: 5177 }
{ Generator: task | monitored: 1 | n_generated: 995 }
\end{CodeOutput}

The simulation has been run for 1000 units of time, and the simulator
has monitored all the state changes and lifetimes of all processes,
which enables any kind of analysis without any additional effort from
the modeller's side. For instance, we may extract a history of the
resource's state to analyse the average number of machines/operatives in
use as well as the average number of jobs/tasks waiting for an
assignment.

\begin{CodeInput}
R> aggregate(cbind(server, queue) ~ resource, get_mon_resources(env), mean)
\end{CodeInput}

\begin{CodeOutput}
   resource   server     queue
1   machine 7.987438 1.0355590
2 operative 3.505732 0.4441298
\end{CodeOutput}

The development of the \pkg{simmer} package started in the second half
of 2014. The initial need for a DES framework for \proglang{R} came up
in projects related to process optimisation in healthcare facilities.
Most of these cases involved patients following a clear trajectory
through a care process. This background is not unimportant, as it lead
to the adoption and implementation of a \emph{trajectory} concept at the
very core of \pkg{simmer}'s DES engine. This strong focus on clearly
defined \emph{trajectories} is somewhat innovative and, more
importantly, very intuitive. Furthermore, this framework relies on a
fast \proglang{C++} simulation core to boost performance and make DES
modelling in \proglang{R} not only effective, but also efficient.

Over time, the \pkg{simmer} package has seen significant improvements
and has been at the forefront of DES for \proglang{R}. Although it is
the most generic DES framework, it is however not the only \proglang{R}
package which delivers such functionality. For example, the \pkg{SpaDES}
package \citep{CRAN:SpaDES} focuses on spatially explicit discrete
models, and the \pkg{queuecomputer} package \citep{CRAN:queuecomputer}
implements an efficient method for simulating queues with arbitrary
arrival and service times. Going beyond the \proglang{R} language, the
direct competitors to \pkg{simmer} are \pkg{SimPy} \citep{SimPy} and
\pkg{SimJulia} \citep{GitHub:SimJulia}, built for respectively the
\proglang{Python} and \proglang{Julia} languages.

\section{The simulation core design}\label{the-simulation-core-design}

The core of any modern discrete-event simulator comprises two main
components: an event list, ordered by time of occurrence, and an event
loop that extracts and executes events. In contrast to other interpreted
languages such as \proglang{Python}, which is compiled by default to an
intermediate byte-code, \proglang{R} code is purely parsed and evaluated
at runtime\footnote{Some effort has been made in this line with the
  \pkg{compiler} package, introduced in \proglang{R} version 2.13.0
  \citep{R:compiler}, furthermore, a JIT-compiler was included in
  \proglang{R} version 3.4.0.}. This fact makes it a particularly slow
language for DES, which consists of executing complex routines (pieces
of code associated to the events) inside a loop while constantly
allocating and deallocating objects (in the event queue).

In fact, first attempts were made in pure \proglang{R} by these authors,
and a minimal process-based implementation with \pkg{R6} classes
\citep{CRAN:R6} proved to be unfeasible in terms of performance compared
to similar approaches in pure \proglang{Python}. For this reason, it was
decided to provide a robust and fast simulation core written in
\proglang{C++}. The \proglang{R} API interfaces with this \proglang{C++}
core by leveraging the \pkg{Rcpp} package
\citep{Eddelbuettel:2011:Rcpp, Eddelbuettel:2013:Rcpp}, which has become
one of the most popular ways of extending \proglang{R} packages with
\proglang{C} or \proglang{C++} code.

The following subsections are devoted to describe the simulation core
architecture. First, we establish the DES terminology used in the rest
of the paper. Then, the architectural choices made are discussed, as
well as the event queue and the \emph{simultaneity problem}, an
important topic that every DES framework has to deal with.

\subsection{Terminology}\label{terminology}

This document uses some DES-specific terminology, e.g., \emph{event},
\emph{state}, \emph{entity}, \emph{process} or \emph{attribute}. Such
standard terms can be easily found in any textbook about DES (refer to
\citet{Banks:2005:Discrete}, for instance). There are, however, some
\pkg{simmer}-specific terms, and some elements that require further
explanation to understand the package architecture.

\begin{description}
\item[Resource]
A passive entity, as it is commonly understood in standard DES
terminology. However, \pkg{simmer} resources are conceived with queuing
systems in mind, and therefore they comprise two internal self-managed
parts:

\begin{description}
\tightlist
\item[Server]
which, conceptually, represents the resource itself. It has a specified
capacity and can be seized and released.
\item[Queue]
A priority queue of a certain size.
\end{description}
\item[Manager]
An active entity, i.e., a process, that has the ability to adjust
properties of a resource (capacity and queue size) at run-time.
\item[Generator]
A process responsible for creating new \emph{arrivals} with a given
interarrival time pattern and inserting them into the simulation model.
\item[Arrival]
A process capable of interacting with resources or other entities of the
simulation model. It may have some attributes and prioritisation values
associated and, in general, a limited lifetime. Upon creation, every
arrival is attached to a given \emph{trajectory}.
\item[Trajectory]
An interlinkage of \emph{activities} constituting a recipe for arrivals
attached to it, i.e., an ordered set of actions that must be executed.
The simulation model is ultimately represented by a set of trajectories.
\item[Activity]
The individual unit of action that allows arrivals to interact with
resources and other entities, perform custom routines while spending
time in the system, move back and forth through the trajectory
dynamically, and much more.
\end{description}

\begin{figure}
\centering
\includegraphics{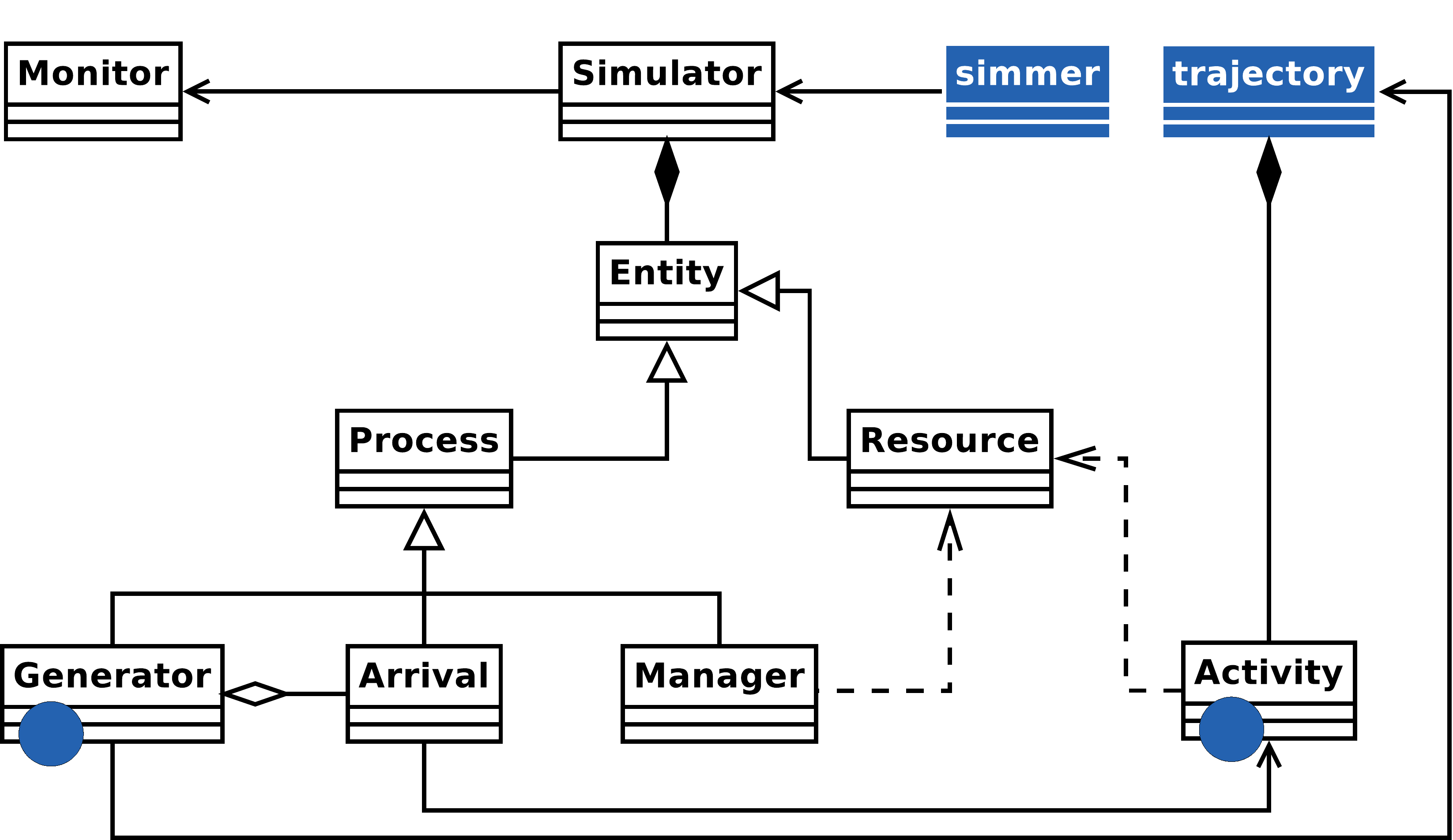}
\caption{UML diagram of the simulation core architecture. Blue classes
represent how \proglang{R} encapsulates the \proglang{C++} core. Blue
circles represent how \proglang{C++} interfaces with
\proglang{R}.\label{architecture}}
\end{figure}

\pagebreak

\subsection{Architecture}\label{architecture}

Extending an \proglang{R} package (or any other piece of software
written in any interpreted language) with compiled code poses an
important trade-off between performance and flexibility: placing too
much functionality into the compiled part produces gains in performance,
but degrades modelling capabilities, and vice versa. The following lines
are devoted to discuss how this trade-off is resolved in \pkg{simmer}.

Figure \ref{architecture} sketches a UML (Unified Modelling Language)
description of the architecture, which constitutes a process-based
design, as in many modern DES frameworks. We draw the attention now to
the \proglang{C++} classes (depicted in white).

The first main component is the \code{Simulator} class. It comprises the
event loop and the event queue, which will be addressed in the next
subsection. The \code{Simulator} provides methods for scheduling and
unscheduling events. Moreover, it is responsible for managing
simulation-wide entities (e.g., resources and generators) and facilities
(e.g., signaling between processes and batches) through diverse
\proglang{C++} unordered maps:

\begin{itemize}
\tightlist
\item
  Maps of resources and processes (generators, arrivals and managers) by
  name.
\item
  A map of pending events, which allows to unschedule a given process.
\item
  Maps of signals subscribed by arrivals and handlers defined for
  different signals.
\item
  Maps for forming batches of arrivals, named and unnamed.
\end{itemize}

This class also holds global attributes and monitoring information.
Thus, monitoring counters, which are derived from the \code{Monitor}
class, are centralised, and they register every change of state produced
during the simulation time. There are five types of built-in changes of
state that are recorded by calling \code{Simulator}'s \code{record\_*()}
methods:

\begin{itemize}
\tightlist
\item
  An arrival is accepted into a resource (served or enqueued). The
  resource notifies about the new status of its internal counters.
\item
  An arrival leaves a resource. The resource notifies the new status of
  its internal counters, and the arrival notifies start, end and
  activity times in that particular resource.
\item
  A resource is modified during runtime (i.e., a change in the capacity
  or queue size). The resource notifies the new status of its internal
  counters.
\item
  An arrival modifies an attribute, one of its own or a global one. The
  arrival notifies the new value.
\item
  An arrival leaves its trajectory by exhausting the activities
  associated (considered as \emph{finished}) or because of another
  reason (\emph{non-finished}, e.g., it is rejected from a resource).
  The arrival notifies global start, end and activity times.
\end{itemize}

As mentioned in the previous subsection, there are two types of
entities: passive ones (\code{Resource}) and active ones (processes
\code{Generator}, \code{Arrival} and \code{Manager}). Generators create
new arrivals, and the latter are the main actors of the simulation
model. Managers can be used for dynamically changing the properties of a
resource (capacity and queue size). All processes share a \code{run()}
method that is invoked by the event loop each time a new event is
extracted from the event list.

There is a fourth kind of process not shown in Figure
\ref{architecture}, called \code{Task}. It is a generic process that
executes a given function once, and it is used by arrivals, resources,
activities and the simulator itself to trigger dynamic actions or split
up events. A \code{Task} is for instance used under the hood to trigger
reneging or to broadcast signals after some delay.

The last main component, completely isolated from the \code{Simulator},
is the \code{Activity} class. This abstract class represents a clonable
object, chainable in a double-linked list to form trajectories. Most of
the activities provided by \pkg{simmer} derive from it. \code{Fork} is
another abstract class (not depicted in Figure \ref{architecture}) which
is derived from \code{Activity}. Any activity supporting the definition
of sub-trajectories must derive from this one instead, such as
\code{Seize}, \code{Branch} or \code{Clone}. All the activities must
implement the virtual methods \code{print()} and \code{run()}.

Finally, it is worth mentioning the couple of blue circles depicted in
Figure \ref{architecture}. They represent the \emph{points of presence}
of \proglang{R} in the \proglang{C++} core, i.e., where the core
interfaces back with \proglang{R} to execute custom user-defined code.

In summary, the \proglang{C++} core is responsible for all the heavy
tasks, i.e., managing the event loop, the event list, generic resources
and processes, collecting all the statistics, and so on. And still, it
provides enough flexibility to the user for modelling the interarrival
times from \proglang{R} and execute any custom user-defined code through
the activities.

\subsection{The event queue}\label{the-event-queue}

The event queue is the most fundamental part of any DES software. It is
responsible for maintaining a list of events to be executed by the event
loop in an ordered fashion by time of occurrence. This last requirement
establishes the need for a data structure with a low access, search,
insertion and deletion complexity. A binary tree is a well-known data
structure that satisfies these properties, and it is commonly used for
this purpose. Unfortunately, binary trees, or equivalent structures,
cannot be efficiently implemented without pointers, and this is the main
reason why pure \proglang{R} is very inefficient for DES.

In \pkg{simmer}, the event queue is defined as a \proglang{C++}
multiset, a kind of associative container implemented as a balanced tree
internally. Apart from the efficiency, it was selected to support event
unscheduling through iterators. Each event holds a pointer to a process,
which will be retrieved and run in the event loop. Events are inserted
in the event queue ordered by 1) time of occurrence and 2) priority.
This secondary order criterion is devoted to solve a common issue for
DES software called \emph{the simultaneity problem}.

\subsubsection{The simultaneity problem}\label{the-simultaneity-problem}

As noted by \citet{Ronngren:1999:EOP:301429.301456},
\citet{Jha:2000:SEL:361026.361032}, there are many circumstances from
which simultaneous events (i.e., events with the same timestamp) may
arise. How they are handled by a DES framework has critical implications
on reproducibility and simulation correctness.

As an example of the implications, let us consider an arrival seizing a
resource at time \(t_{i-1}\), which has \code{capacity=1} and
\code{queue_size=0}. At time \(t_{i}\), two simultaneous events happen:
1) the resource is released, and 2) another arrival tries to seize the
resource. It is indisputable what should happen in this situation: the
new arrival seizes the resource while the other continues its path. But
note that if 2) is executed \emph{before} 1), the new arrival is
rejected (!). Therefore, it is obvious that release events must always
be executed \emph{before} seize events.

If we consider a dynamically managed resource (i.e., its capacity
changes over time) and, instead of the event 1) in the previous example,
the manager increases the capacity of the resource, we are in the very
same situation. Again, it is obvious that resource managers must be
executed \emph{before} seize attempts.

A further analysis reveals that, in order to preserve correctness and
prevent a simulation crash, it is necessary to break down resource
releases in two parts with different priorities: the release in itself
and a post-release event that tries to serve another arrival from the
queue. Thus, every resource manager must be executed \emph{after}
releases and \emph{before} post-releases. This and other issues are
solved with a priority system (see Table \ref{priorities}) embedded in
the event list implementation that provides a deterministic and
consistent execution of simultaneous events.

\begin{longtable}[]{@{}ll@{}}
\toprule
Priority & Event\tabularnewline
\midrule
\endfirsthead
\toprule
Priority & Event\tabularnewline
\midrule
\endhead
\texttt{PRIORITY\_MAX} & Modify a generator (e.g., activate or
deactivate it)\tabularnewline
\texttt{PRIORITY\_RELEASE} & Resource release\tabularnewline
\texttt{PRIORITY\_MANAGER} & Manager action (e.g., resource capacity
change)\tabularnewline
\texttt{PRIORITY\_RELEASE\_POST} & Resource post-release (i.e., serve
from the queue)\tabularnewline
\texttt{PRIORITY\_GENERATOR} & Generate new arrivals\tabularnewline
\ldots{} & General activities\tabularnewline
\texttt{PRIORITY\_MIN} & Other tasks (e.g., a timer for
reneging)\tabularnewline
\bottomrule
\caption{Priority system (in decreasing order) and events
associated.\label{priorities}}
\end{longtable}

\section[The simmer API]{The \pkg{simmer} API}

The \proglang{R} API exposed by \pkg{simmer} comprises two main
elements: the \code{simmer} environment (or \emph{simulation
environment}) and the \code{trajectory} object, which are depicted in
Figure \ref{architecture} (blue classes). As we will see throughout this
section, simulating with \pkg{simmer} simply consists of building a
simulation environment and one or more trajectories. For this purpose,
the API is composed of verbs and actions that can be chained together.
For easy-of-use, these have been made fully compatible with the pipe
operator (\code{\%>\%}) from the \pkg{magrittr} package.

\subsection{The trajectory object}\label{the-trajectory-object}

A \emph{trajectory} can be defined as a recipe and consists of an
ordered set of \emph{activities}. The idea behind this concept is very
similar to the idea behind \pkg{dplyr} for data manipulation
\citep{CRAN:dplyr}. To borrow the words of H. Wickham, ``by constraining
your options, it simplifies how you can think about'' discrete-event
modelling. Activities are \emph{verbs} that correspond to common
functional DES blocks.

The \code{trajectory()} method instantiates the object, and activities
can be appended using the \code{\%>\%} operator:

\begin{CodeInput}
R> traj0 <- trajectory() %>%
R+   log_("Entering the trajectory") %>%
R+   timeout(10) %>%
R+   log_("Leaving the trajectory")
\end{CodeInput}

The trajectory above illustrates the two most basic activities
available: displaying a message (\code{log_()}) and spending some time
in the system (\code{timeout()}). An arrival attached to this trajectory
will execute the activities in the given order, i.e., it will display
``Entering the trajectory'', then it will spend 10 units of (simulated)
time, and finally it will display ``Leaving the trajectory''.

The example uses \emph{fixed parameters}: a string and a numeric value
respectively. However, at least the main parameter for all activities
(this is specified in the documentation) can also be what we will call a
\emph{dynamical parameter}, i.e., a function. This thus, although not
quite useful yet, is also valid:

\begin{CodeInput}
R> traj1 <- trajectory() %>%
R+   log_(function() "Entering the trajectory") %>%
R+   timeout(function() 10) %>%
R+   log_(function() "Leaving the trajectory")
\end{CodeInput}

Also, trajectories can be split apart, joined together and modified:

\begin{CodeInput}
R> traj2 <- join(traj0[c(1, 3)], traj0[2])
R> traj2[1] <- traj2[3]
R> traj2
\end{CodeInput}

\begin{CodeOutput}
trajectory: anonymous, 3 activities
{ Activity: Timeout      | delay: 10 }
{ Activity: Log          | message }
{ Activity: Timeout      | delay: 10 }
\end{CodeOutput}

There are many activities available. We will briefly review them by
categorising them into different topics.

\subsubsection{Arrival properties}\label{arrival-properties}

Arrivals are able to store attributes and modify these using
\code{set\_attribute()}. Attributes consist of pairs \code{(key, value)}
(character and numeric respectively) which by default are set \emph{per
arrival} unless they are defined as \code{global}. As we said before,
all activities support at least one dynamical parameter. In the case of
\code{set\_attribute()}, this is the \code{value} parameter.

Attributes can be retrieved in any \proglang{R} function by calling
\code{get\_attribute()}, whose first argument must be a \code{simmer}
object. For instance, the following trajectory prints \code{81}:

\begin{CodeInput}
R> env <- simmer()
R>
R> traj <- trajectory() %>%
R+   set_attribute("weight", 80) %>%
R+   set_attribute("weight", function() get_attribute(env, "weight") + 1) %>%
R+   log_(function() paste0("My weight is ", get_attribute(env, "weight")))
\end{CodeInput}

\pagebreak

Arrivals also hold a set of three prioritisation values for accessing
resources:

\begin{description}
\tightlist
\item[\code{priority}]
A higher value equals higher priority. The default value is the minimum
priority, which is 0.
\item[\code{preemptible}]
If a preemptive resource is seized, this parameter establishes the
minimum incoming priority that can preempt this arrival (the activity is
interrupted and another arrival with a \code{priority} greater than
\code{preemptible} gains the resource). In any case, \code{preemptible}
must be equal or greater than \code{priority}, and thus only higher
priority arrivals can trigger preemption.
\item[\code{restart}]
Whether the ongoing activity must be restarted after being preempted.
\end{description}

These three values are established for all the arrivals created by a
particular generator, but they can also be dynamically changed on a
per-arrival basis using the \code{set\_prioritization()} and
\code{get\_prioritization()} activities, in the same way as attributes.

\subsubsection{Interaction with
resources}\label{interaction-with-resources}

The two main activities for interacting with resources are
\code{seize()} and \code{release()}. In their most basic usage, they
seize/release a given \code{amount} of a resource specified by name. It
is also possible to change the properties of the resource with
\code{set\_capacity()} and \code{set\_queue\_size()}.

The \code{seize()} activity is special in the sense that the outcome
depends on the state of the resource. The arrival may successfully seize
the resource and continue its path, but it may also be enqueued or
rejected and dropped from the trajectory. To handle these special cases
with total flexibility, \code{seize()} supports the specification of two
optional sub-trajectories: \code{post.seize}, which is followed after a
successful seize, and \code{reject}, followed if the arrival is
rejected. As in every activity supporting the definition of
sub-trajectories, there is a boolean parameter called \code{continue}.
For each sub-trajectory, it controls whether arrivals should continue to
the activity following the \code{seize()} in the main trajectory after
executing the sub-trajectory.

\begin{CodeInput}
R> patient <- trajectory() %>%
R+   log_("arriving...") %>%
R+   seize(
R+     "doctor", 1, continue = c(TRUE, FALSE),
R+     post.seize = trajectory("accepted patient") %>%
R+       log_("doctor seized"),
R+     reject = trajectory("rejected patient") %>%
R+       log_("rejected!") %>%
R+       seize("nurse", 1) %>%
R+       log_("nurse seized") %>%
R+       timeout(2) %>%
R+       release("nurse", 1) %>%
R+       log_("nurse released")
R+   ) %>%
R+   timeout(5) %>%
R+   release("doctor", 1) %>%
R+   log_("doctor released")
R>
R> env <- simmer() %>%
R+   add_resource("doctor", capacity = 1, queue_size = 0) %>%
R+   add_resource("nurse", capacity = 10, queue_size = 0) %>%
R+   add_generator("patient", patient, at(0, 1)) %>%
R+   run()
\end{CodeInput}

\begin{CodeOutput}
0: patient0: arriving...
0: patient0: doctor seized
1: patient1: arriving...
1: patient1: rejected!
1: patient1: nurse seized
3: patient1: nurse released
5: patient0: doctor released
\end{CodeOutput}

The value supplied to all these methods may be a dynamical parameter. On
the other hand, the resource name must be fixed. There is a special
mechanism to select resources dynamically: the \code{select()} activity.
It marks a resource as selected for an arrival executing this activity
given a set of \code{resources} and a \code{policy}. There are several
policies implemented internally that can be accessed by name:

\begin{description}
\tightlist
\item[\code{shortest-queue}]
The resource with the shortest queue is selected.
\item[\code{round-robin}]
Resources will be selected in a cyclical nature.
\item[\code{first-available}]
The first available resource is selected.
\item[\code{random}]
A resource is randomly selected.
\end{description}

Its \code{resources} parameter is allowed to be dynamical, and there is
also the possibility of defining custom policies. Once a resource is
selected, there are special versions of the aforementioned activities
for interacting with resources without specifying its name, such as
\code{seize\_selected()}, \code{set\_capacity\_selected()} and so on.

\subsubsection{Interaction with
generators}\label{interaction-with-generators}

There are four activities specifically intended to modify generators. An
arrival may \code{activate()} or \code{deactivate()} a generator, but
also modify with \code{set\_trajectory()} the trajectory to which it
attaches the arrivals created, or set a new interarrival distribution
with \code{set\_distribution()}. For dynamically selecting a generator,
the parameter that specifies the generator name in all these methods can
be dynamical.

\begin{CodeInput}
R> traj <- trajectory() %>%
R+   deactivate("dummy") %>%
R+   timeout(1) %>%
R+   activate("dummy")
R>
R> simmer() %>%
R+   add_generator("dummy", traj, function() 1) %>%
R+   run(10) %>%
R+   get_mon_arrivals()
\end{CodeInput}

\begin{CodeOutput}
    name start_time end_time activity_time finished replication
1 dummy0          1        2             1     TRUE           1
2 dummy1          3        4             1     TRUE           1
3 dummy2          5        6             1     TRUE           1
4 dummy3          7        8             1     TRUE           1
\end{CodeOutput}

\subsubsection{Branching}\label{branching}

A branch is a point in a trajectory in which one or more
sub-trajectories may be followed. Two types of branching are supported
in \pkg{simmer}. The \code{branch()} activity places the arrival in one
of the sub-trajectories depending on some condition evaluated in a
dynamical parameter called \code{option}. It is the equivalent of an
\code{if/else} in programming, i.e., if the value of \code{option} is
\(i\), the \(i\)-th sub-trajectory will be executed. On the other hand,
the \code{clone()} activity is a \emph{parallel} branch. It does not
take any option, but replicates the arrival \code{n-1} times and places
each one of them into the \code{n} sub-trajectories supplied.

\begin{CodeInput}
R> env <- simmer()
R>
R> traj <- trajectory() %>%
R+   branch(
R+     option = function() round(now(env)), continue = c(FALSE, TRUE),
R+     trajectory() %>% log_("branch 1"),
R+     trajectory() %>% log_("branch 2")
R+   ) %>%
R+   clone(
R+     n = 2,
R+     trajectory() %>% log_("clone 0"),
R+     trajectory() %>% log_("clone 1")
R+   ) %>%
R+   synchronize(wait = TRUE) %>%
R+   log_("out")
R>
R> env %>%
R+   add_generator("dummy", traj, at(1, 2)) %>%
R+   run() %>% invisible
\end{CodeInput}

\begin{CodeOutput}
1: dummy0: branch 1
2: dummy1: branch 2
2: dummy1: clone 0
2: dummy1: clone 1
2: dummy1: out
\end{CodeOutput}

Note that \code{clone()} is the only exception among all activities
supporting sub-trajectories that does not accept a \code{continue}
parameter. By default, all the clones continue in the main trajectory
after this activity. To remove all of them except for one, the
\code{synchronize()} activity may be used.

\subsubsection{Loops}\label{loops}

There is a mechanism, \code{rollback()}, for going back in a trajectory
and thus executing loops over a number of activities. This activity
causes the arrival to step back a given \code{amount} of activities
(that can be dynamical) a number of \code{times}. If a \code{check}
function returning a boolean is supplied, the \code{times} parameter is
ignored and the arrival determines whether it must step back each time
it hits the \code{rollback}.

\begin{CodeInput}
R> hello <- trajectory() %>%
R+   log_("Hello!") %>%
R+   timeout(1) %>%
R+   rollback(amount = 2, times = 2)
R>
R> simmer() %>%
R+   add_generator("hello_sayer", hello, at(0)) %>%
R+   run() %>% invisible
\end{CodeInput}

\begin{CodeOutput}
0: hello_sayer0: Hello!
1: hello_sayer0: Hello!
2: hello_sayer0: Hello!
\end{CodeOutput}

\subsubsection{Batching}\label{batching}

Batching consists of collecting a number of arrivals before they can
continue their path in the trajectory as a unit\footnote{A concrete
  example of this is the case where a number of people (the arrivals)
  together take, or rather seize, an elevator (the resource).}. This
means that if, for instance, 10 arrivals in a batch try to seize a unit
of a certain resource, only one unit may be seized, not 10. A batch may
be splitted with \code{separate()}, unless it is marked as
\code{permanent}.

\begin{CodeInput}
R> roller <- trajectory() %>%
R+   batch(10, timeout = 5, permanent = FALSE) %>%
R+   seize("rollercoaster", 1) %>%
R+   timeout(5) %>%
R+   release("rollercoaster", 1) %>%
R+   separate()
\end{CodeInput}

By default, all the arrivals reaching a batch are joined into it, and
batches wait until the specified number of arrivals are collected.
Nonetheless, arrivals can avoid joining the batch under any constraint
if an optional function returning a boolean, \code{rule}, is supplied.
Also, a batch may be triggered before collecting a given amount of
arrivals if some \code{timeout} is specified. Note that batches are
shared only by arrivals directly attached to the same trajectory.
Whenever a globally shared batch is needed, a common \code{name} must be
specified.

\pagebreak

\subsubsection{Asynchronous programming}\label{asynchronous-programming}

There are a number of methods enabling asynchronous events. The
\code{send()} activity broadcasts one or more \code{signals} to all the
arrivals subscribed to them. Signals can be triggered immediately or
after some \code{delay}. In this case, both parameters, \code{signals}
and \code{delay}, can be dynamical. Arrivals are able to block and
\code{wait()} until a certain signal is received.

Arrivals can subscribe to \code{signals} and (optionally) assign a
\code{handler} using the \code{trap()} activity. Upon a signal
reception, the arrival stops the current activity and executes the
\code{handler}\footnote{The \code{handler} parameter accepts a
  trajectory object. Once the handler gets called, it will route the
  arrival to this sub-trajectory.} if provided. Then, the execution
returns to the activity following the point of interruption.
Nonetheless, trapped signals are ignored when the arrival is waiting in
a resource's queue. The same applies inside a batch: all the signals
subscribed before entering the batch are ignored. Finally, the
\code{untrap()} activity can be used to unsubscribe from \code{signals}.

\begin{CodeInput}
R> t_blocked <- trajectory() %>%
R+   trap(
R+     "you shall pass",
R+     handler = trajectory() %>%
R+       log_("got a signal!")
R+   ) %>%
R+   log_("waiting...") %>%
R+   wait() %>%
R+   log_("continuing!")
R>
R> t_signal <- trajectory() %>%
R+   log_("you shall pass") %>%
R+   send("you shall pass")
R>
R> simmer() %>%
R+   add_generator("blocked", t_blocked, at(0)) %>%
R+   add_generator("signaler", t_signal, at(5)) %>%
R+   run() %>% invisible
\end{CodeInput}

\begin{CodeOutput}
0: blocked0: waiting...
5: signaler0: you shall pass
5: blocked0: got a signal!
5: blocked0: continuing!
\end{CodeOutput}

By default, signal handlers may be interrupted as well by other signals,
meaning that a \code{handler} may keep restarting if there are frequent
enough signals being broadcasted. If an uninterruptible \code{handler}
is needed, this can be achieved by setting the flag \code{interruptible}
to \code{FALSE} in \code{trap()}.

\subsubsection{Reneging}\label{reneging}

Besides being rejected while trying to seize a resource, arrivals are
also able to leave the trajectory at any moment, synchronously or
asynchronously. Namely, reneging means that an arrival abandons the
trajectory at a given moment. The most simple activity enabling this is
\code{leave}, which immediately triggers the action given some
probability. Furthermore, \code{renege\_in()} and \code{renege\_if()}
trigger reneging asynchronously after some timeout \code{t} or if a
\code{signal} is received respectively, unless the action is aborted
with \code{renege\_abort()}. Both \code{renege\_in()} and
\code{renege\_if()} accept an optional sub-trajectory, \code{out}, that
is executed right before leaving.

\begin{CodeInput}
R> bank <- trajectory() %>%
R+   log_("Here I am") %>%
R+   renege_in(
R+     5,
R+     out = trajectory() %>%
R+       log_("Lost my patience. Reneging...")
R+   ) %>%
R+   seize("clerk", 1) %>%
R+   renege_abort() %>%
R+   log_("I'm being attended") %>%
R+   timeout(10) %>%
R+   release("clerk", 1) %>%
R+   log_("Finished")
R>
R> simmer() %>%
R+   add_resource("clerk", 1) %>%
R+   add_generator("customer", bank, at(0, 1)) %>%
R+   run() %>% invisible
\end{CodeInput}

\begin{CodeOutput}
0: customer0: Here I am
0: customer0: I'm being attended
1: customer1: Here I am
6: customer1: Lost my patience. Reneging...
10: customer0: Finished
\end{CodeOutput}

\subsection{The simulation
environment}\label{the-simulation-environment}

The simulation environment manages resources and generators, and
controls the simulation execution. The \code{simmer()} method
instantiates the object, after which resources and generators can be
appended using the \code{\%>\%} operator:

\begin{CodeInput}
R> env <- simmer()
R>
R> env %>%
R+   add_resource("res_name", 1) %>%
R+   add_generator("arrival", traj0, function() 25)
\end{CodeInput}

\begin{CodeOutput}
simmer environment: anonymous | now: 0 | next: 0
{ Resource: res_name | monitored: TRUE | server status: 0(1) | queue... }
{ Generator: arrival | monitored: 1 | n_generated: 0 }
\end{CodeOutput}

Then, the simulation can be executed, or \code{run()}, \code{until} a
stop time:

\begin{CodeInput}
R> env %>%
R+   run(until=40)
\end{CodeInput}

\begin{CodeOutput}
25: arrival0: Entering the trajectory
35: arrival0: Leaving the trajectory
\end{CodeOutput}

\begin{CodeOutput}
simmer environment: anonymous | now: 40 | next: 50
{ Resource: res_name | monitored: TRUE | server status: 0(1) | queue... }
{ Generator: arrival | monitored: 1 | n_generated: 2 }
\end{CodeOutput}

There are a number of methods for extracting information, such as the
simulation time (\code{now()}), future scheduled events (\code{peek()}),
and \emph{getters} for obtaining resources' and generators' parameters
(capacity, queue size, server count and queue count; number of arrivals
generated so far). There are also several \emph{setters} available for
resources and generators (capacity, queue size; trajectory,
distribution).

A \code{simmer} object can be \code{reset()} and re-run. However, there
is a special method, \code{wrap()}, intended to extract all the
information from the \proglang{C++} object encapsulated into a
\code{simmer} environment and to deallocate that object. Thus, most of
the \emph{getters} work also when applied to wrapped environments, but
such an object cannot be reset or re-run anymore.

\subsubsection{Resources}\label{resources}

A \pkg{simmer} resource, as stated in Section \ref{terminology},
comprises two internal self-managed parts: a server and a priority
queue. Three main parameters define a resource: \code{name} of the
resource, \code{capacity} of the server and \code{queue_size} (0 means
no queue). Resources are monitored and non-preemptive by default.
Preemption means that if a high priority arrival becomes eligible for
processing, the resource will temporarily stop the processing of one (or
more) of the lower priority arrivals being served. For preemptive
resources, the \code{preempt_order} defines which arrival should be
stopped first if there are many lower priority arrivals, and it assumes
a first-in-first-out (FIFO) policy by default. Any preempted arrival is
enqueued in a dedicated queue that has a higher priority over the main
one (i.e., it is served first). The \code{queue_size_strict} parameter
controls whether this dedicated queue must be taken into account for the
queue size limit, if any. If this parameter enforces the limit, then
rejection may occur in the main queue.

\subsubsection{Generators}\label{generators}

Three main parameters define a generator: a \code{name_prefix} for each
generated arrival, a trajectory to attach them to and an interarrival
\code{distribution}. Parameters \code{priority}, \code{preemptible} and
\code{restart} have been described in Section \ref{arrival-properties}.
The monitoring flag accepts several levels in this case:

\begin{enumerate}
\def\labelenumi{\arabic{enumi}.}
\setcounter{enumi}{-1}
\tightlist
\item
  No monitoring enabled.
\item
  Arrival monitoring.
\item
  Level 1 \(+\) attribute monitoring.
\end{enumerate}

The interarrival \code{distribution} must return one or more
interarrival times for each call. Internally, generators create as many
arrivals as values returned by this function. They do so with zero-delay
and re-schedule themselves with a delay equal to the sum of the values
obtained. Whenever a negative interarrival value is obtained, the
generator stops.

\subsection{Monitoring and data
retrieval}\label{monitoring-and-data-retrieval}

There are three methods for obtaining monitored data (if any) about
arrivals, resources and attributes. They can be applied to a single
simulation environment or to a list of environments, and the returning
object is always a data frame, even if no data was found. Each processed
simulation environment is treated as a different replication, and a
numeric column named \code{replication} is added to every returned data
frame with environment indexes as values.

\begin{description}
\tightlist
\item[\code{get\_mon\_arrivals()}]
Returns timing information per arrival: \code{name} of the arrival,
\code{start_time}, \code{end_time}, \code{activity_time} (time not spent
in resource queues) and a flag, \code{finished}, that indicates whether
the arrival exhausted its activities (or was rejected). By default, this
information is referred to the arrivals' entire lifetime, but it may be
obtained on a per-resource basis by specifying \code{per_resource=TRUE}.
\item[\code{get\_mon\_resources()}]
Returns state changes in resources: \code{resource} name, \code{time}
instant of the event that triggered the state change, \code{server}
count, \code{queue} count, \code{capacity}, \code{queue_size},
\code{system} count (\code{server} \(+\) \code{queue}) and system
\code{limit} (\code{capacity} \(+\) \code{queue_size}).
\item[\code{get\_mon\_attributes()}]
Returns state changes in attributes: \code{name} of the attribute,
\code{time} instant of the event that triggered the state change, name
of \code{key} that identifies the attribute and \code{value}.
\end{description}

\section[Modelling with simmer]{Modelling with \pkg{simmer}}

The following subsections aim to provide some basic modelling examples.
The topics addressed are queuing systems, replication, parallelisation
and some best practices. We invite the reader to learn about a broader
selection of activities and modelling techniques available in the
package vignettes, which cover the use of attributes, loops, batching,
branching, shared events, reneging and advanced uses of resources among
others.

\subsection{Queuing systems}\label{queuing-systems}

The concept of \emph{trajectory} developed in \pkg{simmer} emerges as a
natural way to simulate a wide range of problems related to
Continuous-Time Markov Chains (CTMC), and more specifically to the
so-called birth-death processes and queuing systems. Indeed,
\pkg{simmer} not only provides very flexible resources (with or without
queue), branches, delays and arrival generators, but they are bundled in
a very comprehensive framework of verbs that can be chained with the
pipe operator. Let us explore the expressiveness of a \pkg{simmer}
trajectory using a \emph{traditional} queuing example: the M/M/1. The
package vignettes include further examples on M/M/c/k systems, queueing
networks and CTMC models.

In Kendall's notation \citep{Kendall:1953}, an M/M/1 system has
exponential arrivals (\textbf{M}/M/1), a single server (M/M/\textbf{1})
with exponential service time (M/\textbf{M}/1) and an infinite queue
(implicit M/M/1/\textbf{\(\infty\)}). For instance, people arriving at
an ATM at rate \(\lambda\), waiting their turn in the street and
withdrawing money at rate \(\mu\). These are the basic parameters of the
system, whenever \(\rho < 1\):

\begin{align}
\rho &= \frac{\lambda}{\mu} &&\equiv \mbox{Server utilisation} \\
N &= \frac{\rho}{1-\rho} &&\equiv \mbox{Average number of customers in the system (queue $+$ server)} \label{eq:N}\\
T &= \frac{N}{\lambda} &&\equiv \mbox{Average time in the system (queue $+$ server) [Little's law]}
\end{align}

If \(\rho \ge 1\), it means that the system is unstable: there are more
arrivals than the server is capable of handling and the queue will grow
indefinitely. The simulation of an M/M/1 system is quite simple using
\pkg{simmer}:

\begin{CodeInput}
R> library(simmer)
R>
R> set.seed(1234)
R>
R> lambda <- 2
R> mu <- 4
R> rho <- lambda/mu
R>
R> mm1.traj <- trajectory() %>%
R+   seize("mm1.resource", amount=1) %>%
R+   timeout(function() rexp(1, mu)) %>%
R+   release("mm1.resource", amount=1)
R>
R> mm1.env <- simmer() %>%
R+   add_resource("mm1.resource", capacity=1, queue_size=Inf) %>%
R+   add_generator("arrival", mm1.traj, function() rexp(1, lambda)) %>%
R+   run(until=2000)
\end{CodeInput}

After the parameter setup, the first code block defines the trajectory:
each arrival will seize the resource, wait some exponential random time
(service time) and release the resource. The second code block
instantiates the simulation environment, creates the resource, attaches
an exponential generator to the trajectory and runs the simulation for
2000 units of time. Note that trajectories can be defined independently
of the simulation environment, but it is recommended to instantiate the
latter in the first place, so that trajectories are able to extract
information from it (e.g., the simulation time).

As a next step, we could extract the monitoring information and perform
some analyses. The extension package \pkg{simmer.plot}
\citep{CRAN:simmer.plot} provides convenience plotting methods to, for
instance, quickly visualise the usage of a resource over time. Figure
\ref{mm1-plot} gives a glimpse of this simulation using this package. In
particular, it shows that the average number of customers in the system
converges to the theoretical value given by Equation \eqref{eq:N}.

\begin{figure}

{\centering \subfloat[Instantaneous usage.\label{fig:mm1-plot1}]{\includegraphics[width=.49\linewidth]{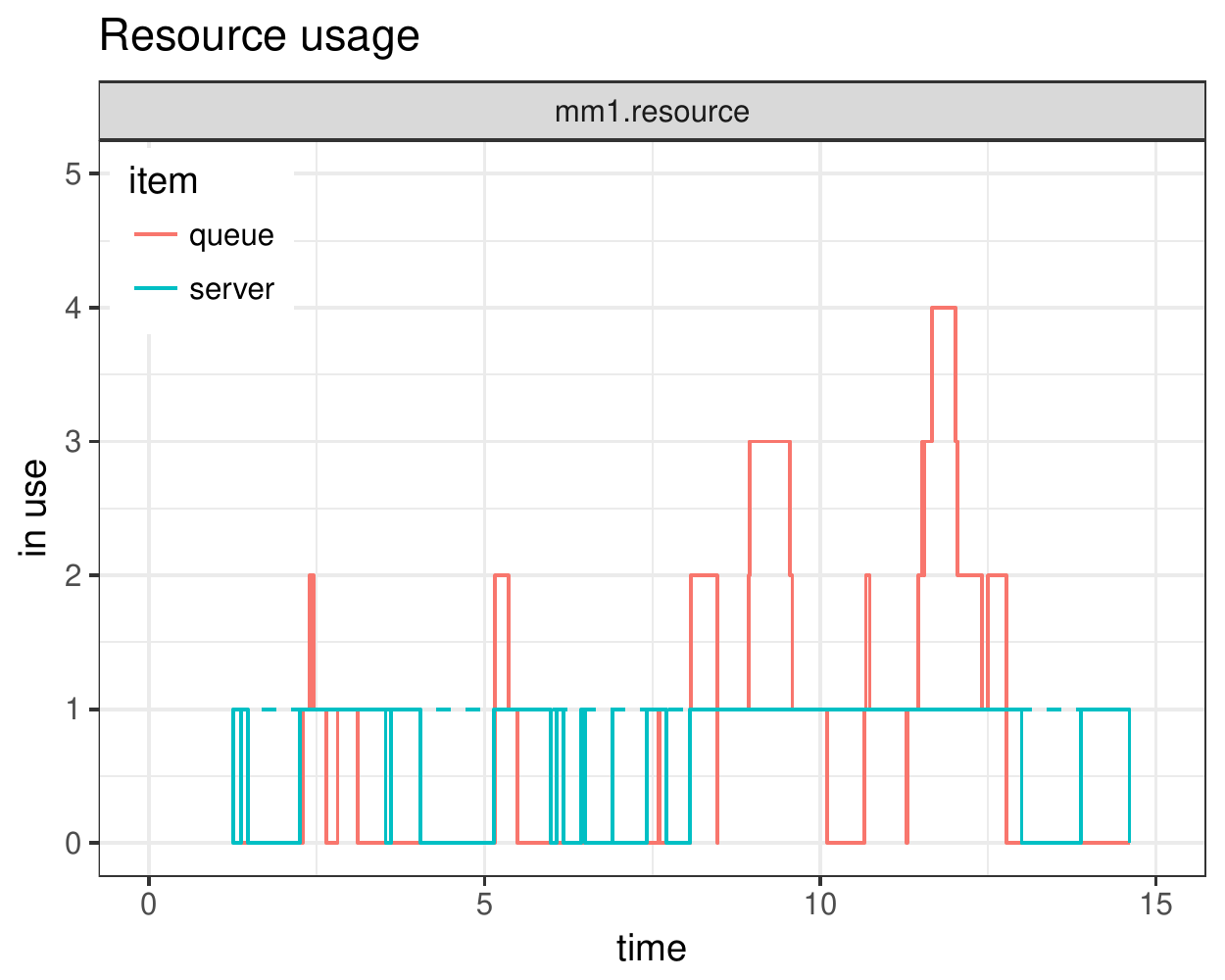} }\subfloat[Convergence over time.\label{fig:mm1-plot2}]{\includegraphics[width=.49\linewidth]{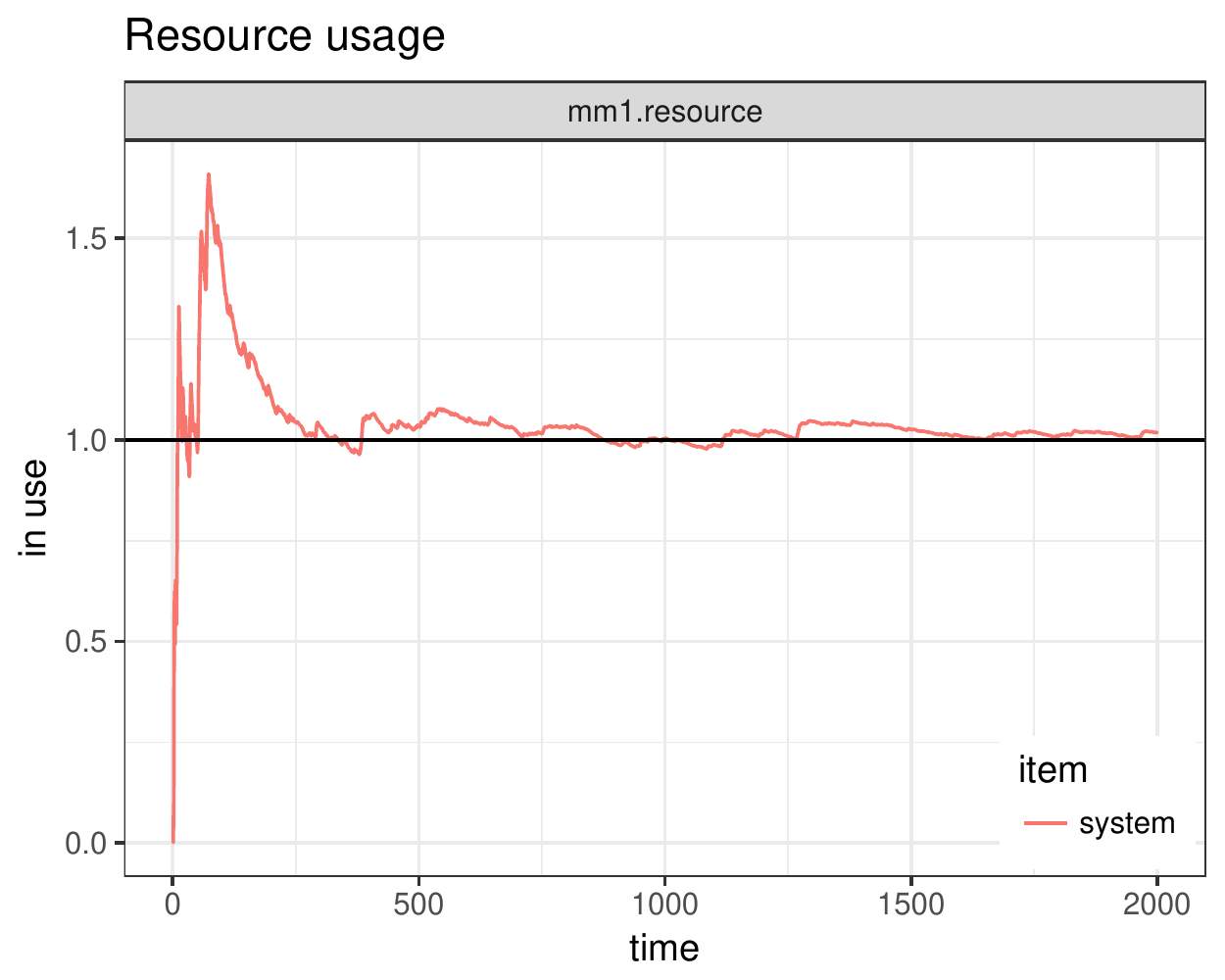} }

}

\caption[Detail of the resource usage\label{mm1-plot}]{Detail of the resource usage\label{mm1-plot}.}\label{fig:mm1-plot}
\end{figure}

\pagebreak

\subsection{Replication and
parallelisation}\label{replication-and-parallelisation}

Typically, running a certain simulation only once is useless. In
general, we will be interested in replicating the model execution many
times, maybe with different initial conditions, and then perform some
statistical analysis over the output. This can be easily achieved using
standard \proglang{R} tools, e.g., \code{lapply()} or similar functions.

Additionally, we can leverage the parallelised version of
\code{lapply()}, \code{mclapply()}, provided by the \pkg{parallel}
package, to speed up this process. Unfortunately, parallelisation has
the shortcoming that we lose the underlying \proglang{C++} objects when
each thread finishes. To avoid losing the monitored data, the
\code{wrap()} method can be used to extract and wrap these data into a
pure \proglang{R} object before the \proglang{C++} object is
garbage-collected.

The following example uses \code{mclapply()} and \code{wrap()} to
perform 100 replicas of the M/M/1 simulation from the previous section
(note that the trajectory is not redefined):

\begin{CodeInput}
R> library(simmer)
R> library(parallel)
R>
R> set.seed(1234)
R>
R> mm1.envs <- mclapply(1:100, function(i) {
R+   simmer() %>%
R+     add_resource("mm1.resource", capacity=1, queue_size=Inf) %>%
R+     add_generator("arrival", mm1.traj, function() rexp(100, lambda)) %>%
R+     run(until=1000/lambda) %>%
R+     wrap()
R+ }, mc.set.seed=FALSE)
\end{CodeInput}

\pagebreak

With all these replicas, we could, for instance, perform a t-test over
\(N\), the average number of customers in the system:

\begin{CodeInput}
R> mm1.data <-
R+   get_mon_arrivals(mm1.envs) %>%
R+   dplyr::group_by(replication) %>%
R+   dplyr::summarise(mean = mean(end_time - start_time))
R>
R> t.test(mm1.data[["mean"]])
\end{CodeInput}

\begin{CodeOutput}

    One Sample t-test

data:  mm1.data[["mean"]]
t = 94.883, df = 99, p-value < 2.2e-16
alternative hypothesis: true mean is not equal to 0
95 percent confidence interval:
 0.4925143 0.5135535
sample estimates:
mean of x
0.5030339
\end{CodeOutput}

\subsection{Best practices}\label{best-practices}

DES modelling can be done in an event-by-event basis, but this approach
is fairly tedious and mostly unpractical. Instead, modern
process-oriented approaches commonly relate to the identification of
resources and processes in a given problem, and the interactions between
them. The \pkg{simmer} package internally follows this paradigm and
exposes generic resources and processes (\emph{arrivals}, in
\pkg{simmer} terminology), so that the user can implement all the
interactions as action sequences (\emph{trajectories}).

There are usually multiple valid ways of mapping the identified
resources and processes into the elements exposed by the \pkg{simmer}
API. For example, let us suppose that we would like to model an alarm
clock beeping every second. In this case, the beep may be identified as
a process, so that we have \emph{different} beeps (multiple arrivals)
entering a \code{beep} trajectory once per second:

\begin{CodeInput}
R> beep <- trajectory() %>%
R+   log_("beeeep!")
R>
R> env <- simmer() %>%
R+   add_generator("beep", beep, function() 1) %>%
R+   run(2.5)
\end{CodeInput}

\begin{CodeOutput}
1: beep0: beeeep!
2: beep1: beeeep!
\end{CodeOutput}

But instead, identifying the alarm clock as the process is equally
valid, and then we have \emph{a single alarm} (single arrival) producing
all the beeps in a loop:

\begin{CodeInput}
R> alarm <- trajectory() %>%
R+   timeout(1) %>%
R+   log_("beeeep!") %>%
R+   rollback(2)
R>
R> env <- simmer() %>%
R+   add_generator("alarm", alarm, at(0)) %>%
R+   run(2.5)
\end{CodeInput}

\begin{CodeOutput}
1: alarm0: beeeep!
2: alarm0: beeeep!
\end{CodeOutput}

These are two common design patterns in \pkg{simmer} for which the
outcome is the same, although there are subtle differences that depend
on the problem being considered and the monitoring requirements.
Furthermore, as a model becomes more complex and detailed, the resulting
mapping and syntax may become more artificious. These issues are shared
in different ways by other frameworks as well, such as \pkg{SimPy}, and
arise due to their generic nature.

Furthermore, the piping mechanism used in the \pkg{simmer} API may
invite the user to produce large monolithic trajectories. However, it
should be noted that it is usually better to break them down into small
manageable pieces. For instance, the following example parametrises the
access to a \code{resource}, where \code{G} refers to arbitrary service
times, and \emph{n} servers are seized. Then, it is used to instantiate
the trajectory shown in the former M/M/1 example:

\begin{CodeInput}
R> xgn <- function(resource, G, n)
R+   trajectory() %>%
R+     seize(resource, n) %>%
R+     timeout(G) %>%
R+     release(resource, n)
R>
R> (mm1.traj <- xgn("mm1.resource", function() rexp(1, mu), 1))
\end{CodeInput}

\begin{CodeOutput}
trajectory: anonymous, 3 activities
{ Activity: Seize        | resource: mm1.resource, amount: 1 }
{ Activity: Timeout      | delay: 0x556861e81830 }
{ Activity: Release      | resource: mm1.resource, amount: 1 }
\end{CodeOutput}

Standard \proglang{R} tools (\code{lapply()} and the like) may also be
used to generate large lists of trajectories with some variations. These
small pieces can be concatenated together into longer trajectories using
\code{join()}, but at the same time, they allow for multiple points of
attachment of arrivals.

During a simulation, trajectories can interact with the simulation
environment in order to extract or modify parameters of interest such as
the current simulation time, attributes, status of resources (get the
number of arrivals in a resource, get or set resources' capacity or
queue size), or status of generators (get the number of generated
arrivals, set generators' attached trajectory or distribution). The only
requirement is that the simulation object must be defined in the same
\proglang{R} environment (or a parent one) \emph{before} the simulation
is started. Effectively, it is enough to detach the \code{run()} method
from the instantiation (\code{simmer()}), namely, they should not be
called in the same pipe. But, for the sake of consistency, it is a good
coding practice to instantiate the simulation object always in the first
place as follows:

\begin{CodeInput}
R> set.seed(1234)
R> env <- simmer()
R>
R> traj <- trajectory() %>%
R+   log_(function() paste0("Current simulation time: ", now(env)))
R>
R> env <- env %>%
R+   add_generator("dummy", traj, at(rexp(1, 1))) %>%
R+   run()
\end{CodeInput}

\begin{CodeOutput}
2.50176: dummy0: Current simulation time: 2.50175860496223
\end{CodeOutput}

\section{Performance evaluation}\label{performance-evaluation}

This section investigates the performance of \pkg{simmer} with the aim
of assessing its usability as a general-purpose DES framework. A first
subsection is devoted to measuring the simulation time of a simple model
relative to \pkg{SimPy} and \pkg{SimJulia}. The reader may find
interesting to compare the expressiveness of each framework. Last but
not least, the final subsection explores the cost of calling
\proglang{R} from \proglang{C++}, revealing the existent trade-off,
inherent to the design of this package, between performance and model
complexity.

All the subsequent tests were performed under Fedora Linux 25 running on
an Intel Core2 Quad CPU Q8400, with \proglang{R} 3.3.3,
\proglang{Python} 2.7.13, \pkg{SimPy} 3.0.9, \proglang{Julia} 0.5.1 and
\pkg{SimJulia} 0.3.14 installed from the default repositories. Absolute
execution times presented here are specific to this platform and
configuration, and thus they should not be taken as representative for
any other system. Instead, the relative performance should be
approximately constant across different systems.

\subsection{Comparison with similar
frameworks}\label{comparison-with-similar-frameworks}

A significant effort has been put into the design of \pkg{simmer} in
order to make it performant enough to run general and relatively large
simulation models in a reasonable amount of time. In this regard, a
relevant comparison can be made against other general-purpose DES
frameworks such as \pkg{SimPy} and \pkg{SimJulia}. To this effect, we
retake the M/M/1 example from Section \ref{queuing-systems}, which can
be bundled into the following test:

\begin{CodeInput}
R> library(simmer)
R>
R> test_mm1_simmer <- function(n, m, mon=FALSE) {
R>   mm1 <- trajectory() %>%
R>     seize("server", 1) %>%
R>     timeout(function() rexp(1, 1.1)) %>%
R>     release("server", 1)
R>
R>   env <- simmer() %>%
R>     add_resource("server", 1, mon=mon) %>%
R>     add_generator("customer", mm1, function() rexp(m, 1), mon=mon) %>%
R>     run(until=n)
R> }
\end{CodeInput}

With the selected arrival rate, \(\lambda=1\), this test simulates an
average of \code{n} arrivals entering a nearly saturated system
(\(\rho=1/1.1\)). Given that \pkg{simmer} generators are able to create
arrivals in batches (i.e., more than one arrival for each function call)
for improved performance, the parameter \code{m} controls the size of
the batch. Finally, the \code{mon} flag enables or disables monitoring.

Let us build now the equivalent model using \pkg{SimPy}, with base
\proglang{Python} for random number generation. We prepare the
\proglang{Python} benchmark from \proglang{R} using the \pkg{rPython}
package \citep{CRAN:rPython} as follows:

\begin{CodeInput}
R> rPython::python.exec("
R> import simpy, random, time
R>
R> def test_mm1(n):
R>   def exp_source(env, lambd, server, mu):
R>       while True:
R>           dt = random.expovariate(lambd)
R>           yield env.timeout(dt)
R>           env.process(customer(env, server, mu))
R>
R>   def customer(env, server, mu):
R>       with server.request() as req:
R>           yield req
R>           dt = random.expovariate(mu)
R>           yield env.timeout(dt)
R>
R>   env = simpy.Environment()
R>   server = simpy.Resource(env, capacity=1)
R>   env.process(exp_source(env, 1, server, 1.1))
R>   env.run(until=n)
R>
R> def benchmark(n, times):
R>   results = []
R>   for i in range(0, times):
R>     start = time.time()
R>     test_mm1(n)
R>     results.append(time.time() - start)
R>   return results
R> ")
\end{CodeInput}

Equivalently, this can be done for \proglang{Julia} and \pkg{SimJulia}
using the \pkg{rjulia} package \citep{GitHub:rJulia}. Once more,
\code{n} controls the number of arrivals simulated on average:

\begin{CodeInput}
R> rjulia::julia_init()
R> rjulia::julia_void_eval("
R> using SimJulia, Distributions
R>
R> function test_mm1(n::Float64)
R>   function exp_source(env::Environment, lambd::Float64,
R>                       server::Resource, mu::Float64)
R>     while true
R>       dt = rand(Exponential(1/lambd))
R>       yield(Timeout(env, dt))
R>       Process(env, customer, server, mu)
R>     end
R>   end
R>
R>   function customer(env::Environment, server::Resource, mu::Float64)
R>     yield(Request(server))
R>     dt = rand(Exponential(1/mu))
R>     yield(Timeout(env, dt))
R>     yield(Release(server))
R>   end
R>
R>   env = Environment()
R>   server = Resource(env, 1)
R>   Process(env, exp_source, 1.0, server, 1.1)
R>   run(env, n)
R> end
R>
R> function benchmark(n::Float64, times::Int)
R>   results = Float64[]
R>   test_mm1(n)
R>   for i = 1:times
R>     push!(results, @elapsed test_mm1(n))
R>   end
R>   return(results)
R> end
R> ")
\end{CodeInput}

It can be noted that in both cases there is no monitoring involved,
because either \pkg{SimPy} nor \pkg{SimJulia} provide automatic
monitoring as \pkg{simmer} does. Furthermore, the resulting code for
\pkg{simmer} is more concise and expressive than the equivalent ones for
\pkg{SimPy} and \pkg{SimJulia}, which are very similar.

\pagebreak

We obtain the reference benchmark with \code{n=1e4} and 20 replicas for
both packages as follows:

\begin{CodeInput}
R> n <- 1e4L
R> times <- 20
R>
R> ref <- data.frame(
R>   SimPy = rPython::python.call("benchmark", n, times),
R>   SimJulia = rjulia::j2r(paste0("benchmark(", n, ".0, ", times, ")"))
R> )
\end{CodeInput}

As a matter of fact, we also tested a small DES skeleton in pure
\proglang{R} provided in \citep[\emph{7.8.3 Extended Example:
Discrete-Event Simulation in \proglang{R}}]{Matloff:2011:ARP:2090080}.
This code was formalised into an \proglang{R} package called \pkg{DES},
available on GitHub\footnote{\url{https://github.com/matloff/des}} since
2014. The original code implemented the event queue as an ordered vector
which was updated by performing a binary search. Thus, the execution
time of this version was two orders of magnitude slower than the other
frameworks. The most recent version on GitHub (as of 2017) takes another
clever approach though: it supposes that the event vector will be short
and approximately ordered; therefore, the event vector is not sorted
anymore, and the next event is found using a simple linear search. These
assumptions hold for many cases, and particularly for this M/M/1
scenario. As a result, the performance of this model is only \(\sim2.2\)
times slower than \pkg{SimPy}. Still, it is clear that pure \proglang{R}
cannot compete with other languages in discrete-event simulation, and
\pkg{DES} is not considered in our comparisons hereafter.

\begin{figure}

{\centering \subfloat[Boxplots for 20 runs of the M/M/1 test with \code{n=1e4}.\label{performance-mm1-init}\label{fig:performance-mm1-plot1}]{\includegraphics[width=.49\linewidth]{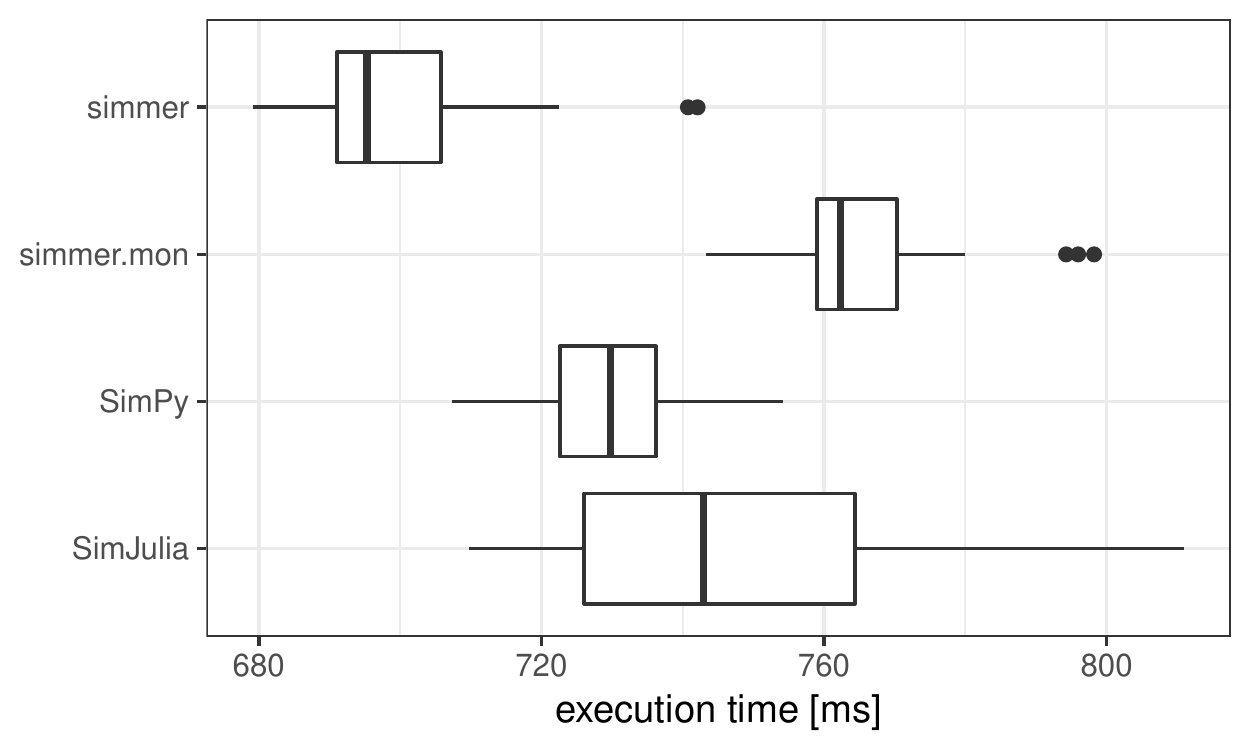} }\subfloat[Performance evolution with the batch size $m$.\label{performance-mm1-plot}\label{fig:performance-mm1-plot2}]{\includegraphics[width=.49\linewidth]{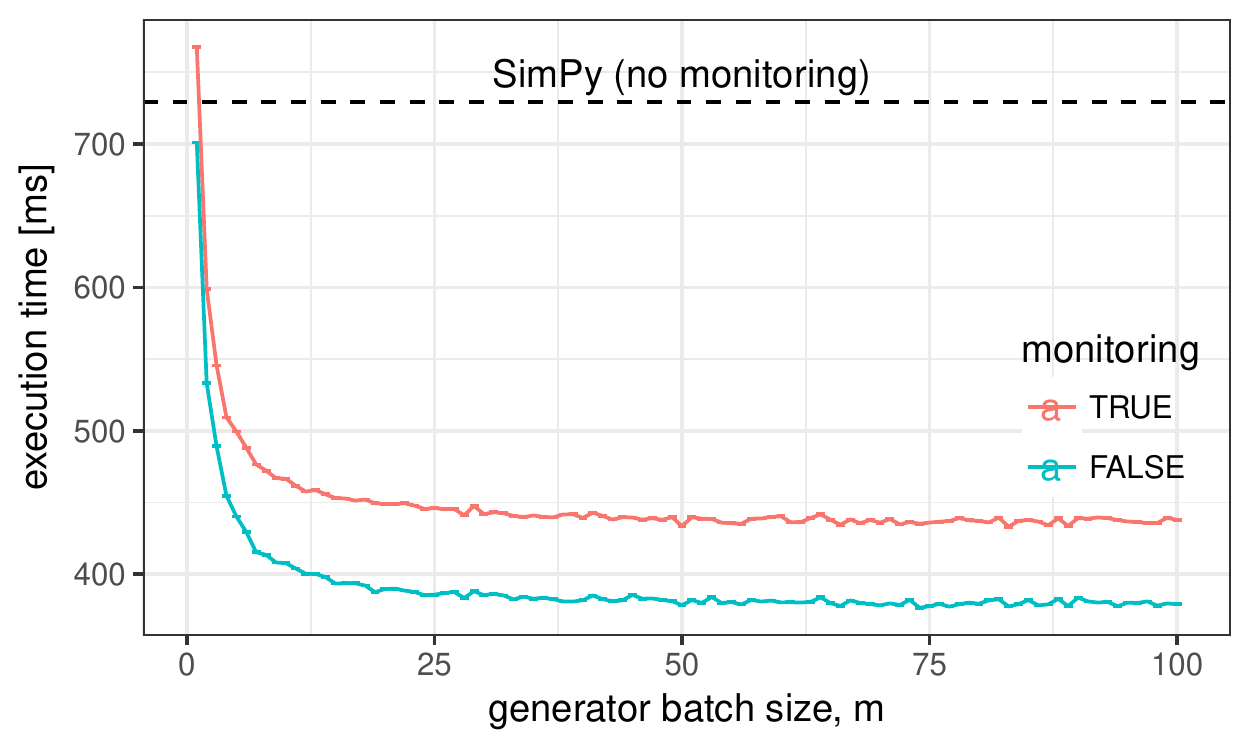} }

}

\caption[Performance comparison]{Performance comparison.}\label{fig:performance-mm1-plot}
\end{figure}

Finally, we set a benchmark for \pkg{simmer} using \pkg{microbenchmark},
again with \code{n=1e4} and 20 replicas for each test. Figure
\ref{performance-mm1-init} shows the output of this benchmark.
\pkg{simmer} is tested both in monitored and in non-monitored mode. The
results show that the performance of \pkg{simmer} is equivalent to
\pkg{SimPy} and \pkg{SimJulia}. The non-monitored \pkg{simmer} shows a
slightly better performance than these frameworks, while the monitored
\pkg{simmer} shows a slightly worse performance.

At this point, it is worth highlighting \pkg{simmer}'s ability to
generate arrivals in batches (hence parameter \(m\)). To better
understand the impact of batched arrival generation, the benchmark was
repeated over a range of \(m\) values \((1, \ldots, 100)\). The results
of the batched arrival generation runs are shown in Figure
\ref{performance-mm1-plot}. This plot depicts the average execution time
of the \pkg{simmer} model with (red) and without (blue) monitoring as a
function of the generator batch size \(m\). The black dashed line sets
the average execution time of the \pkg{SimPy} model to serve as a
reference.

The performance with \code{m=1} corresponds to what has been shown in
Figure \ref{performance-mm1-init}. But as \(m\) increases, \pkg{simmer}
performance quickly improves and becomes \(\sim1.6\) to \(1.9\) times
faster than \pkg{SimPy}. Surprisingly, there is no additional gain with
batches greater than 40-50 arrivals at a time, but there is no penalty
either with bigger batches. Therefore, it is always recommended to
generate arrivals in big batches whenever possible.

\subsection[The cost of calling R from C++]{The cost of calling \proglang{R} from \proglang{C++}}

The \proglang{C++} simulation core provided by \pkg{simmer} is quite
fast, as we have demonstrated, but performance is adversely affected by
numerous calls to \proglang{R}. The practice of calling \proglang{R}
from \proglang{C++} is generally strongly discouraged due to the
overhead involved. However, in the case of \pkg{simmer}, it not only
makes sense, but is even fundamental in order to provide the user with
enough flexibility to build all kinds of simulation models.
Nevertheless, this cost must be known, and taken into account whenever a
higher performance is needed.

To explore the cost of calling \proglang{R} from \proglang{C++}, let us
define the following test:

\begin{CodeInput}
R> library(simmer)
R>
R> test_simmer <- function(n, delay) {
R+   test <- trajectory() %>%
R+     timeout(delay)
R+
R+   env <- simmer() %>%
R+     add_generator("test", test, at(1:n)) %>%
R+     run(Inf)
R+
R+   arrivals <- get_mon_arrivals(env)
R+ }
\end{CodeInput}

This toy example performs a very simple simulation in which \code{n}
arrivals are attached (in one shot, thanks to the convenience function
\code{at()}) to a \code{test} trajectory at \(t=1, 2, ..., n\). The
trajectory consists of a single activity: a timeout with some
configurable \code{delay} that may be a fixed value or a function call.
Finally, after the simulation, the monitored data is extracted from the
simulation core to \proglang{R}. Effectively, this is equivalent to
generating a data frame of \code{n} rows (see the example output in
Table \ref{test-simmer-output-table}).

\begin{longtable}[]{@{}lrrrlr@{}}
\toprule
Name & Start time & End time & Activity time & Finished &
Replication\tabularnewline
\midrule
\endfirsthead
\toprule
Name & Start time & End time & Activity time & Finished &
Replication\tabularnewline
\midrule
\endhead
test0 & 1 & 2 & 1 & TRUE & 1\tabularnewline
test1 & 2 & 3 & 1 & TRUE & 1\tabularnewline
test2 & 3 & 4 & 1 & TRUE & 1\tabularnewline
\bottomrule
\caption{Output from the \code{test\_simmer()}
function.\label{test-simmer-output-table}}
\end{longtable}

As a matter of comparison, the following \code{test_R_for()} function
produces the very same data using base \proglang{R}:

\begin{CodeInput}
R> test_R_for <- function(n) {
R>   name <- character(n)
R>   start_time <- numeric(n)
R>   end_time <- numeric(n)
R>   activity_time <- logical(n)
R>   finished <- numeric(n)
R>
R>   for (i in 1:n) {
R>     name[i] <- paste0("test", i-1)
R>     start_time[i] <- i
R>     end_time[i] <- i+1
R>     activity_time[i] <- 1
R>     finished[i] <- TRUE
R>   }
R>
R>   arrivals <- data.frame(
R>     name=name,
R>     start_time=start_time,
R>     end_time=end_time,
R>     activity_time=activity_time,
R>     finished=finished,
R>     replication = 1
R>   )
R> }
\end{CodeInput}

Note that we are using a \code{for} loop to mimic the behaviour of
\pkg{simmer}'s internals, of how monitoring is made, but we concede the
advantage of pre-allocated vectors to \proglang{R}. A second base
\proglang{R} implementation, which builts upon the \texttt{lapply()}
function, is implemented as the \code{test_R_lapply()} function:

\begin{CodeInput}
R> test_R_lapply <- function(n) {
R>   as.data.frame(do.call(rbind, lapply(1:n, function(i) {
R>     list(
R>       name = paste0("test", i - 1),
R>       start_time = i,
R>       end_time = i + 1,
R>       activity_time = 1,
R>       finished = TRUE,
R>       replication = 1
R>     )
R>   })))
R> }
\end{CodeInput}

The \code{test_simmer()}, \code{test_R_for()} and \code{test_R_lapply()}
functions all produce exactly the same data in a similar manner (cfr.
Table \ref{test-simmer-output-table}). Now, we want to compare how a
delay consisting of a function call instead of a fixed value impacts the
performance of \pkg{simmer}, and we use \code{test_R_for()} and
\code{test_R_lapply()} as yardsticks.

To this end, the \pkg{microbenchmark} package
\citep{CRAN:microbenchmark} is used. The benchmark was executed with
\code{n=1e5} and 20 replicas for each test. Table
\ref{performance-table} shows a summary of the resulting timings. As we
can see, \pkg{simmer} is \(\sim4.4\) times faster than \code{for}-based
base \proglang{R} and \(\sim3.6\) times faster than \code{lapply}-based
base \proglang{R} on average when we set a fixed delay. On the ther
hand, if we replace it for a function call, the execution becomes
\(\sim6.5\) times slower, or \(\sim1.5\) times slower than
\code{for}-based base \proglang{R}. It is indeed a quite good result if
we take into account the fact that base \proglang{R} pre-allocates
memory, and that \pkg{simmer} is doing a lot more internally. But still,
these results highlight the overheads involved and encourage the use of
fixed values instead of function calls whenever possible.

\begin{longtable}[]{@{}lrrrr@{}}
\toprule
Expr & Min & Mean & Median & Max\tabularnewline
\midrule
\endfirsthead
\toprule
Expr & Min & Mean & Median & Max\tabularnewline
\midrule
\endhead
test\_simmer(n, 1) & 429.8663 & 492.365 & 480.5408 &
599.3547\tabularnewline
test\_simmer(n, function() 1) & 3067.9957 & 3176.963 & 3165.6859 &
3434.7979\tabularnewline
test\_R\_for(n) & 2053.0840 & 2176.164 & 2102.5848 &
2438.6836\tabularnewline
test\_R\_lapply(n) & 1525.6682 & 1754.028 & 1757.7566 &
2002.6634\tabularnewline
\bottomrule
\caption{Execution time
(milliseconds).\label{performance-table}}
\end{longtable}

\section{Summary}\label{summary}

The \pkg{simmer} package presented in this paper brings a generic yet
powerful process-oriented Discrete-Event Simulation framework to
\proglang{R}. \pkg{simmer} combines a robust and fast simulation core
written in \proglang{C++} with a rich and flexible \proglang{R} API. The
main modelling component is the \emph{activity}. Activities are chained
together with the pipe operator into \emph{trajectories}, which are
common paths for processes of the same type. \pkg{simmer} provides a
broad set of activities, and allows the user to extend their
capabilities with custom \proglang{R} functions.

Monitoring is automatically performed by the underlying simulation core,
thereby enabling the user to focus on problem modelling. \pkg{simmer}
enables simple replication and parallelisation with standard
\proglang{R} tools. Data can be extracted into \proglang{R} data frames
from a single simulation environment or a list of environments, each of
which is marked as a different replication for further analysis.

Despite the drawbacks of combining \proglang{R} calls into
\proglang{C++} code, \pkg{simmer} shows a good performance combined with
high flexibility. It is currently one of the most extensive DES
frameworks for \proglang{R} and provides a mature workflow for truly
integrating DES into \proglang{R} processes.

\section*{Acknowledgements}

We thank the editors and the anonymous referee for their thorough
reviews and valuable comments, which have been of great help in
improving this paper. Likewise, we thank Norman Matloff for his advice
and support. Last but not least, we are very grateful for vignette
contributions by Duncan Garmonsway, and for all the fruitful ideas for
new or extended features by several users via the \code{simmer-devel}
mailing list and GitHub.

\renewcommand\refname{References}
\bibliography{jss.bib}

\end{document}